\documentclass[11pt,a4paper,leqno]{article}

\usepackage[all]{xy}

\usepackage{amsfonts}
\usepackage{amsthm}

\def\R{\mathbb R}

\def\dom{\mathrm {dom}\,}
\def\LL{\mathrm L}

\newtheorem{teor2}{Theorem}

\newtheorem{prop2}{Proposition}

\newtheorem{obs2}{Remark}

\newcommand{\ddd}{:=}

\begin{document}

 \title{Mathematical justification of the Aharonov-Bohm hamiltonian}
\author{C\'esar R. de Oliveira {\small and} Marciano Pereira\thanks{On
leave of absence from Universidade Estadual de
Ponta Grossa, PR, Brazil.}\\
\vspace{-0.6cm}
\small
\it Departamento de Matem\'{a}tica -- UFSCar, \small \it S\~{a}o Carlos,
SP, 13560-970 Brazil\\ \\}
\date{\today}

\maketitle

\begin{abstract} It is presented, in the framework of nonrelativistic
quantum mechanics, a justification of the usual
Aharonov-Bohm hamiltonian (with solenoid of radius greater than zero).
This is obtained by way of increasing sequences
of finitely long solenoids together with a natural impermeability
procedure; further, both limits commute. Such
rigorous limits are in the  strong resolvent sense and in both $\R^2$ and
$\R^3$ spaces.
\end{abstract}

\vspace{1cm}



\vspace{1cm}

Given a cylindrical current-carrying solenoid $\mathcal S$ of infinite
length and radius $a>0$, centered at the origin
and axis in the
$z$ direction, there is a constant magnetic field $\mathbf B=(0,0,B)$
confined in ${\mathcal S}^\circ$, the interior
of $\mathcal S$, and vanishing in its exterior region $\mathcal S'$. The
solenoid is considered impermeable
(impenetrable), in the sense that the motion of a spinless particle (of mass
$m=1/2$ and electric charge $q$) outside the solenoid has no contact with
its interior, particularly with the magnetic
field
$\mathbf B$. If $\mathbf A$ is the vector potential generating this
magnetic field, that is, $\mathbf B= \nabla
\times\mathbf A$, the usual hamiltonian operator describing the quantum
motion of this charged particle is given by
(with $\hbar=1$)
\[ H_{\mathrm{AB}} = \left(\mathbf p - \frac qc \mathbf A\right)^2,\quad
\mathbf p = -i\nabla,
\] with Dirichlet boundary conditions, i.e., the functions $\psi$ in the
domain of $H_{\mathrm{ AB}}$ are supposed to
vanish $\psi=0$ at the solenoid boundary (the precise domain of
$H_{\mathrm{AB}}$ is described just before
Proposition~\ref{propLfixo}). Observable effects, as wavefunction phase
differences, are predicted and confirmed in
many experiments, even though the particle is confined to
$\mathcal S'$ (see the original paper
\cite{AB1959} and
\cite{PT,OlPo,Ruijs,MVG,Weder,Eskin} for  detailed descriptions and a long
list of additional references). Since the
vector potential is not assumed to (identically) vanish in the exterior
region $\mathcal S'$, the interpretation in
\cite{AB1959}, and followed by a huge amount of papers, is that $\mathbf
A$ plays a prominent role in quantum
mechanics, so that these measurable effects would be regard to be caused
exclusively to $\mathbf A$, and not just to
the magnetic field $\mathbf B$. Since then, this has been called {\it the
Aharonov-Bohm effect} (in spite of such
question had been considered previously \cite{Franz,EhSi}) and it is
directly related to the acceptance of
$H_{\mathrm{AB}}$ above as the quantum model of such situation
(particularly, the presence of the vector potential in
the hamiltonian).

Note that usually the papers devoted to the Aharonov-Bohm effect in
different contexts simply accept the above
hamiltonian operator $H_{\mathrm{AB}}$ prescription (suitably adapted;
e.g., two solenoids) and interpretations. The
goal of this communication is to comment on the difficulties in the
quantization process in this setting and, mainly,
to give grounds for $H_{\mathrm{AB}}$ from a combination of physical
modeling and precise mathematical arguments.
There are other (although related) versions of the Aharonov-Bohm effect
(e.g., with electric potentials), but the
above is the most considered and traditional one; furthermore, many works
consider the idealized case of a solenoid of
radius zero (for instance
\cite{AB1959,Ruijs,DStovc,AdamiTeta}, to mention a few), but here we
concentrate on the more realistic case of radius
$a>0$.

There are controversies over the interpretation of $\mathbf A$ as a real
physical variable, that is, mistrusts of the
existence of the Aharonov-Bohm effect as stated above. For instance, that
the  phase difference could be eliminated by
using gauge transformations
\cite{BoLoi,Boetal}; explanation via the hydrodynamical viewpoint in
quantum mechanics \cite{CG}, whose equations admit
a solution where the vector potential appears explicitly, and such
solution corresponds to a hamiltonian
 with the vector potential included; some authors argue that the
experimental results could be explained by a border
effect and the magnetic field (also due to poor solenoid impermeability)
in a region accessible to the electric
particles
\cite{Roy, HS}---see comments and critiques in
\cite{Greenberger,Klein,Lipkin}.

The acceptance of $H_\mathrm{AB}$, with the explicit appearance of nonzero
$\mathbf A$ outside the impermeable
solenoid, even though $\mathbf B= \nabla \times \mathbf A=0$ there, is
primarily based on an application of Stokes
theorem: if
$\mathcal C$ is a (closed) loop in $\mathcal S'$ around the solenoid,
enclosing an area $\mathcal A$, then it is
assumed that the total magnetic flux crossing $\mathcal A$ is
\[
\Phi = \int_{\mathcal A} \mathbf B\cdot d\mathbf a = \oint_{\mathcal
C}\mathbf A\cdot d\mathbf l,
\]and it is argued that  a phase difference should be expected between
paths from the left and right pieces of
$\mathcal C$ (if $\alpha=q\Phi/(2\pi c)$ is not an integer number).
However, this argument presents mathematical and
physical difficulties that should be carefully justified. From the
mathematical point of view it involves  Stokes
theorem in a multiply connected domain and its application is not
guaranteed (although it is if $\mathcal A$ does not
intersect the interior ${\mathcal S}^\circ$); from the physical point of
view the assumed (electromagnetic)
impermeability of the solenoid should, in principle, inhibit also any
nonzero vector potential in
$\mathcal S'$ from sources in ${\mathcal S}^\circ$. In summary, the
acceptance of $H_{\mathrm{AB}}$ involves an
explicitly choice that needs an explanation (without mention the Dirichlet
boundary conditions; see ahead). Actually,
this is a reflection of the fact that  quantization in multiply connected
domains is not a well-posed question (and
here with some structure
$\mathbf B\ne0$ inside the hole!). Clearly, geometrical and topological
aspects have also been invoked to study the
Aharonov-Bohm effect (see, e.g., \cite{Martin,HLS}). The  effect of an electric field induced by a slowly switching on flux inside the solenoid was studied by Weisskopf \cite{Weisskopf} in 1961. 

In what follows we present a justification of the hamiltonian
$H_{\mathrm{AB}}$. We propose to consider first a
solenoid
$\mathcal S_L$ of finite length
$2L>0$ and also permeable. Recall that a current-carrying finite solenoid
generates a nonzero magnetic field in its
exterior and, since it is also considered permeable, Stokes theorem may be
applied; therefore the corresponding
hamiltonian operator is well defined and with no boundary condition at the
solenoid border.  We model the
impermeability by a sequence of positive potentials
$V_n$ which vanish in the solenoid exterior
$\mathcal S_L'$ region and goes to infinity in its interior
${\mathcal S}_L^\circ$ as $n\to\infty$ \cite{MVG} (to the best of authors knowledge the first one to propose the solenoid impermeability via increasing potentials $V_n$ was Kretzschmar \cite{Kretzschmar}). Then we discuss the
limits of solenoid of infinite length, i.e.,
$L\to\infty$, and impermeability
$n\to\infty$ (so getting a multiply connected region) by showing they
exist (in the resolvent sense \cite{RS1}) and,
finally, that both limits commute, that is, it does not matter which limit
is taken first, and the resulting
hamiltonian is always
$H_{\mathrm{AB}}$. Such limits are in the strong resolvent sense in $\R^2$
and $\R^3$ and we discuss both
simultaneously, since the arguments are almost the same.

Few papers have explicitly considered a finite solenoid \cite{Roy,BLoudon}
in this context; also some mathematically
nonrigorous limiting process are discussed in \cite{BerryAB} in order to
justify the hamiltonian. However, the
arguments may not be considered in the typical criteria of rigor of
mathematical physics we demand here, and this is
our main contribution. One difficulty is that the deficiency indices of
the Aharonov-Bohm hamiltonian $H_{\mathrm{AB}}$
with domain
$C_0^\infty(\mathcal S')$ are both infinite, which leads to a plethora of
self-adjoint extensions; all self-adjoint
extensions of this operator will appear elsewhere
\cite{deOPereira}.

It is worth mentioning the experiments conducted  by Tonomura and collaborators \cite{PT} with toroidal magnets, which have the advantage of no magnetic flux with leaks; recently a rigorous approach to the scattering in this case (and more general ones) has been done in \cite{BW}.

Now we go into details of the idea sketched above for the justification of
$H_\mathrm{AB}$. Let $x=(x_1,x_2,x_3)$
denote the cartesian coordinates in
$\R^3$;  the interior of the finite solenoid $\mathcal S_L$, symmetrically
disposed with respect to the plane
$x_1,x_2$, is
\[ {\mathcal S}_L^\circ = \{(x_1,x_2,x_3): x_1^2+x_2^2<a^2,\;|x_3|<L \},
\]and denote by $\chi_L$ its characteristic function, that is,
$\chi_L(x)=1$ if $x\in {\mathcal S}_L^\circ$ and
$\chi_L(x)=0$ otherwise. The sequence of potential barriers will be
$V_n(x) = n\chi_L(x)$. If $\mathbf A_L$ denotes
the vector potential generated by this  finite permeable solenoid, then
the corresponding hamiltonian is ($0<L\le
\infty$;  note that we write
$\mathbf A=\mathbf A_{L=\infty}$ and $\mathcal S=\mathcal S_{L=\infty}$)
\[ H_{L,n} = \left(\mathbf p - \frac qc \mathbf A_L\right)^2+V_n,\quad
\dom H_{L,n} = \mathcal H^2(\R^d),\,d=2,3,
\]where $\mathcal H^2$ denotes the usual Sobolev space domain of the free
hamiltonian (that is, the negative laplacian)
$-
\Delta$.  In case of
$\R^2$ we just restrict the vector potential and $V_n$ to the plane and
$\mathcal S\cap \R^2$ is a disk centered at
the origin. 

In $\R^3$ there is the possibility of the particle running into the finite solenoid at a point with $x_3=\pm L$ (with total area $\alpha_t=2\pi 
a^2$), which is physically different from entering through the lateral border of the solenoid (with area $\alpha_l=2\pi a\times (2L)$), 
but the potential barrier $V_n$ equally hinders the entrance of the particle from any direction. This effect becomes less and less important as $L$ increases, since the area ratio  $\alpha_t/\alpha_l\to0$ as $L\to\infty$ (note also that, in fact, $\alpha_t$ does not depend on $L$) and for large $L$ the solenoid top and bottom will generally be far away from the electron motion; furthermore, this effect is not present in two-dimensions. Hence, it will not be modeled here.

In both dimensions $d=2,3,$ the finiteness of the solenoid and
permeability make the modeling more feasible, and for
each finite-valued pair $n,L$ the hamiltonian $H_{L,n}$ is a well-posed
operator and self-adjoint. Note that
$\mathbf A_L$ is a bounded and continuous vector function; for instance,
in $\R^2$, by using cylindrical coordinates
$(\rho,\phi,z)$, $z=0$, and the calculation in \cite{Jackson} of the
vector potential of a circular current loop, we
find that (in a particular gauge) the $\rho,z$ components of $\mathbf A_L$
vanish, whereas the $\phi$ component depends
only on
$\rho$ and is given by
\[ A_{L,\phi}(\rho)  = \frac{\Phi}{4\pi^2 a} \int_{-L}^L dz'
\int_0^{2\pi}d\phi'\, \frac{\cos \phi'}{(\rho^2 + a^2
+{z'}^2 -2a\rho\cos \phi')^{1/2}}.
\]Now, for $\rho\ne a$ (the solenoid border $\rho=a$ is a set of zero
Lebesgue measure), we have the expected pointwise
convergence of
$\mathbf A_L$ to
$\mathbf A$ as $L\to\infty$ in $\R^2$, whose $\phi$ component of $\mathbf
A$ is well known and given by
$A_{\phi}(\rho)=
\Phi/(2\pi \rho)$ if $a\le \rho$, and $A_{\phi}(\rho)= \Phi \rho/(2\pi
a^2)$ if $0\le \rho\le a$. Similarly for the
pointwise convergence as $L\to\infty$ of vector potentials in $\R^3$. See
the Appendix for details.

In the particular case of an infinite length solenoid $L=\infty$  in
$\R^3$, the  impermeable limit
$n\to\infty$ was considered in \cite{MVG}; by using Kato-Robinson theorem
\cite{Davies} it was shown that
$H_{\infty,n}$ converges to $H_{\mathrm{AB}}$ with domain
\[
\dom H_{\mathrm{AB}}=\mathcal H^2(\mathcal S')\cap \mathcal H^1_0(\mathcal
S')
\] in the strong resolvent sense as $n\to\infty$, and since elements of
$\mathcal H^1_0(\mathcal S')$ vanish at the
solenoid border (in the sense of Sobolev traces), Dirichlet boundary
conditions have showed up in this situation. Since
the same procedure of
\cite{MVG} for impermeability applies to the case of finite solenoids
$\mathcal S_L$ (with $L$ fixed),  we obtain (in
dimensions
$d=2,3$)

\begin{prop2}\label{propLfixo} As $n\to\infty$ the operator sequence
$H_{L,n}$ converges in the strong resolvent sense
to the operator
\[ H_{L,\infty} := \left(\mathbf p - \frac qc \mathbf A_L\right)^2,\quad
\dom H_{L,\infty}= \mathcal H^2(\mathcal
S_L')\cap
\mathcal H^1_0(\mathcal S_L').
\]
\end{prop2}

\

Fix now $n$.  If $\psi\in C_0^\infty(\R^d)$ and $\mathrm{supp}\,\psi$
denotes its support, then
\[
\| H_{L,n}\psi-H_{\infty,n}\psi \|^2 = \int_{\mathrm{supp}\,\psi} \left|2i
(\mathbf A_L-\mathbf A)\cdot\nabla\psi+
(\mathbf A_L^2-\mathbf A^2)\psi
\right|^2dx,
\]and since as $L\to\infty$ we have the pointwise limit $\mathbf A_L\to
\mathbf A$, it follows that
$H_{L,n}\psi\to H_{\infty,n}\psi$ by Lebesgue dominated convergence. Since
the set $C_0^\infty(\R^d)$ is a core of both
$H_{\infty,n}$ and $H_{L,n}$, for all $L>0$, an application of
Theorem~VIII.25 of \cite{RS1} implies

\begin{prop2}\label{propnfixo} For each fixed $n$, the operator sequence
$H_{L,n}$ converges to $H_{\infty,n}$ in the
strong resolvent sense as
$L\to\infty$.
\end{prop2}

Let $R_i(T)=(T-i)^{-1}$ denote  the resolvent  of a self-adjoint operator
$T$ at the complex number $i$. For $\psi\in
\LL^2(\mathcal S')$ we have
\[
\|R_i(H_{L,\infty})\psi - R_i(H_{\mathrm{AB}})\psi\|
\le \|R_i(H_{L,\infty})\psi - R_i(H_{L,n})\psi\|
\]
\[+
\|R_i(H_{L,n})\psi - R_i(H_{\infty,n})\psi\|  + \|R_i(H_{\infty,n})\psi -
R_i(H_{\mathrm{AB}})\psi\|,
\]and, given $\epsilon>0$, by Proposition~\ref{propnfixo}, if $L$ is large
enough we have \[
\|R_i(H_{L,n})\psi - R_i(H_{\infty,n})\psi\|<\epsilon/3,
\] and after fixing such $L$ we subsequently take $n$ large enough so
that, by Proposition~\ref{propLfixo} and the
resolvent convergence
$H_{\infty,n}\to H_{\mathrm{AB}}$ \cite{MVG},
\[
\|R_i(H_{L,\infty})\psi - R_i(H_{L,n})\psi\| <\epsilon/3,\quad
\|R_i(H_{\infty,n})\psi -
R_i(H_{\mathrm{AB}})\psi\|<\epsilon/3,
\]respectively, so that
\[
\|R_i(H_{L,\infty})\psi - R_i(H_{\mathrm{AB}})\psi\|<\epsilon
\] for $L$ large enough. We have proved:

\begin{prop2}\label{propOutroLim} The operator $H_{L,\infty}$ converges to
$H_{\mathrm{AB}}$ in the strong resolvent
sense as $L\to\infty$.
\end{prop2}

Let $P_0$ denote the projection operator $\LL^2(\R^d)\to \LL^2(\mathcal
S')$. If $\psi\in \LL^2(\R^d)$, then
\[
\| R_i(H_{\mathrm{AB}}) P_0\psi-R_i(H_{L,n})P_0\psi\|
\]
\[
\le \| R_i(H_{\mathrm{AB}}) P_0\psi-R_i(H_{L,\infty})P_0\psi\| + \|
R_i(H_{L,\infty}) P_0\psi-R_i(H_{L,n})P_0\psi\|.
\] By the above propositions both terms on the rhs vanish as
$L,n\to\infty$, and so we conclude

\begin{teor2}\label{teorLn}
$H_{L,n}\to H_{\mathrm{AB}}$ in the strong resolvent sense as
$L,n\to\infty$, independently of the way both limits are
taken.
\end{teor2}

See \cite{Davies,MVG} for a discussion of resolvent convergence when the
domain of the limit operator is not dense in
the original space (as $\LL^2(\mathcal S')$ is not dense in
$\LL^2(\R^d)$). Theorem~\ref{teorLn} says that the same
operator
$H_{\mathrm{AB}}$ is obtained independently of the way the limits of
infinitely long solenoid and impermeability are
processed. For instance, both operations could be done simultaneously by
taken, say, $n=L$ and then $L\to\infty$, etc.
In particular, the limits $L\to\infty$ and $n\to\infty$ do commute. This
is summarized in the  diagram ahead.
\[
\xymatrix{  H_{L,n} \ar@{-{>}}[r]^{n\to\infty}  \ar@{-{>}}[d]_{L\to\infty}
\ar@{-{>}}[dr]|{L,n\to\infty} & H_{L,\infty}
\ar@{-{>}}[d]^{L\to\infty}  \\  H_{\infty,n} \ar@{-{>}}[r]_{n\to\infty} &
H_{\mathrm{AB}} \\  }
\]

Therefore, we are justified in using $H_{\mathrm{AB}}$ while modeling an
infinitely long and impermeable solenoid,
even though we are in a situation of multiply connectedness and with a
magnetic field restricted to the (impenetrable)
region.

\begin{obs2} a) For each fixed $n$ it is possible to check that $H_{L,n}$
converges to $H_{\infty,n}$ in the strong
sense in
$\mathcal H^2(\R^d)$ as $L\to\infty$, for $d=2,3$.

b) By using different techniques, for $d=2,3$ it is possible to show that
for each $L<\infty$ fixed, $H_{L,n}$
converges to
$H_{L,\infty}$ in the uniform resolvent sense as $n\to\infty$. This
uniform convergence also holds for
$L=\infty$ in $\R^2$; however, such uniform convergence should not be
expected to occur in
$\R^3$ when $L=\infty$, because the solenoid border is not compact in this
case.
\end{obs2}

The limit procedures discussed here constitute a step further and
complementary to \cite{MVG}, which has considered
only infinitely long solenoids.

It is intriguing that the (usually just formal) convergence of the
limiting processes to $H_{\mathrm AB}$ has led
different  authors to extremely opposite conclusions:  whereas Magni and
Valz-Gris  (\cite{MVG}, pp.\ 185-186)
concluded that ``The way of coming to that hamiltonian, however, makes it
clear that there is no cogent reason to
attribute vector potentials any physical activity...,'' Berry
\cite{BerryAB} argues that such limits justify the
exclusive quantum role of potentials. At least with respect to this work,
we decided to keep back from this
controversy and restrict ourselves to the above diagram.

\section*{Appendix}
In this appendix we find the expression of the vector potential
$\mathbf A_{L}$ generated by a finite solenoid of length $2L$ in
$\R^3$, in a suitable gauge. Then we show
that its pointwise convergence to $\mathbf A$ as $L\to\infty$. Everything works  in the plane $\R^2$. This fact
was used in the proofs above.

\subsubsection*{Vector potential of a finite solenoid}

The starting point is the vector potential due to a circular current loop
performed in~\cite{Jackson}, Section~5.5.
Then an integration over a density of loops gives the desired vector
potential.  Consider a circular loop of radius
$a>0$ centered at (cartesian coordinates) $(0,0,z')$, $z'\geq 0$, and
parallel to the  plane $xy$. Let $\mathbf{x}'$ be
a  point of the loop and $\mathbf{x}$  a general point in $\R^3$, whose
spherical coordinates are
$\mathbf{x}'=(r',\theta',\phi')$ and $\mathbf{x}=(r,\theta,\phi)$,
respectively.

 The only nonzero component of the current density $\mathbf{J}$ is in the
$\phi$ direction  and, by following
\cite{Jackson}, it is given by
$$J_{\phi} =
I\,\delta(\cos\theta'-\displaystyle\frac{z'}{\sqrt{a^2+{z'}^2}})\,\frac{\delta(r'-\sqrt{a^2+{z'}^2})}{\sqrt{a^2+{z'}^2}},$$
with $I$ denoting the loop electric current. Due to the  symmetry of the
problem, it is possible to assume that the
resulting vector potential has only component the $\phi$ direction, which
actually does not depend on $\phi$; then
select
$\phi = 0$ in the computation that follows. One has
$$A^{z'}_{\phi}(r,\theta)=
\frac{I}{c\sqrt{a^2+{z'}^2}}\int{r'^2dr'd\Omega'\,\frac{\cos\phi'\,\delta(\cos\theta'-\frac{z'}{\sqrt{a^2+{z'}^2}})\,\delta(r'-\sqrt{a^2+{z'}^2})}
{|\mathbf{x}-\mathbf{x}'|}},$$
with
$|\mathbf{x}-\mathbf{x}'| = [r^2+r'^2-2rr'(\cos\theta\,
\cos\theta'+\sin\theta\, \sin\theta'\,\cos\phi')]^{1/2}$ and
\linebreak $d\Omega' = \sin\theta' d\theta' d\phi'$.

On integrating with respect to $r' = \sqrt{a^2+{z'}^2}$, and then with
respect to $\theta'$, with $\cos\theta' =
\displaystyle\frac{z'}{r'}$ and $\sin\theta' = \displaystyle\frac{a}{r'}$,
one  finds
\[
A^{z'}_{\phi}(r,\theta) =
\frac{I}{c\sqrt{a^2+{z'}^2}}\int_0^{2\pi}{(a^2+{z'}^2)
\frac{a}{\sqrt{a^2+{z'}^2}}
d\phi'\,\frac{\cos\phi'}{|\mathbf{x}-\mathbf{x}'|}},
\]
that is,
\[
A^{z'}_{\phi}(r,\theta) = \displaystyle\frac{I\,a}{c}\]
\[\times
\int_0^{2\pi}{d\phi'\,\frac{\cos\phi'}{[r^2+a^2+{z'}^2-2r\sqrt{a^2+{z'}^2}(\cos\theta\,\frac{z'}{\sqrt{a^2+{z'}^2}}+
\sin\theta \, \frac{a}{\sqrt{a^2+{z'}^2}}\,\cos\phi')]^{1/2}}}.
\]
Similarly for $z'\leq 0$.

Thus, the vector potential of the finite solenoid of length $2L$ at a
point $\mathbf x=(r,\theta,\phi)$ in spherical
coordinates is
$\mathbf{A}_L = (0,0,A_{L,\phi})$, where $$A_{L,\phi}(r,\theta) = n
\int_{-L}^L dz'\,A^{z'}_{\phi}(r,\theta) =
\frac{\Phi}{4 {\pi}^2 a}\int_{-L}^{L} dz' \int_0^{2\pi}
d\phi'\,\frac{\cos\phi'}{f(r,\theta,z',\phi')},$$
and $n$ is the number of  loops by length unit in the solenoid, $\Phi$ the
magnetic flux  (so  that $\frac{\Phi}{4
{\pi}^2 a}=\frac{n I a}{c}$) and, finally,
\[
f(r,\theta,z',\phi')\ddd (r^2 + a^2 + {z'}^2 - 2rz'\,\cos\theta
-2ra\,\sin\theta\,\cos\phi')^{1/2}.
\]Here we use the notation $A_\phi(r,\theta)=A_{\infty,\phi}(r,\theta)$
for the $\phi$ component of the vector
potential in case $L=\infty$.

Note that for $\theta = \pi/2$ we have $z=0$ and we obtain the vector
potential in a point
$\mathbf{x}=(r,\pi/2,\phi)$ of the $xy$ plane
\[
A_{L,\phi}(r,\frac{\pi}{2}) = \frac{\Phi}{4 {\pi}^2 a}\int_{-L}^{L} dz'
\int_0^{2\pi}{d\phi'\,\frac{\cos\phi'}{(r^2 + a^2 + z'^2 -
2ra\,\cos\phi')^{1/2}}},
\] which in polar coordinates was denoted by $A_{L,\phi}(\rho)$ above. It
can also be expressed in terms of complete
elliptic integrals $K(k)$ e $E(k)$ \cite{BLoudon}, that is,
\[
A_{L,\phi}(\rho) = \frac{\Phi}{{\pi}^2 a}\int_{-L}^{L} dz' \frac{(2-k^2)
K(k)-2 E(k)}{k^2[(a+\rho)^2+z'^2]^{1/2}},
\]
and $k$ is given by $k^2 = 4 a \rho / [(a+\rho)^2+z'^2]$.

\

\subsubsection*{Convergence as $L\to\infty$}
Fix $r,\theta$. In three situations the term
\[
\left|\frac{2 r a
\sin\theta\cos\phi'}{r^2+a^2+z'^2-2 r z' \cos\theta}\right|
\] is uniformly small:  either 1)~$r\ll a$, or 2)~$r\gg a$ and $r\gg 1$ or
3)~for large $z'\gg1$ and $z'\gg a$. In any
of such situations one has
$$\frac{\cos \phi'}{f(r,\theta,z',\phi')}= \frac{\cos \phi'}{g(r,\theta,z')^{1/2}}+ \frac{r
a\,\sin\theta\cos^2\phi'}{g(r,\theta,z')^{3/2}}+\frac{3}{2}\frac{(r a
\sin\theta)^2\cos^3\phi'}{g(r,\theta,z')^{5/2}}+O(r^{-4},z'^{-7}),$$
with $g(r,\theta,z')\ddd r^2 + a^2 + z'^2 - 2rz'\,\cos\theta$.
Note that the integrals of the first and third terms on the rhs above vanish. Then, the error in the approximation of
$A_\phi(r,\theta)$ by
$A_{L,\phi}(r,\theta)$ can be estimated by (for $L$ large)
\[
\left|A_\phi(r,\theta)-A_{L,\phi}(r,\theta)\right|= \left|\left(\int_{L}^\infty + \int_{-\infty}^{-L} \right)
dz'\,A^{z'}_{\phi}(r,\theta)\right| 
\]
\[
=\left|\left(\int_{L}^\infty + \int_{-\infty}^{-L} \right) dz' \int_0^{2\pi}
d\phi'\,\frac{\cos\phi'}{f(r,\theta,z',\phi')}\right|
\]
\[
\le \mathrm{cte} \left|\int_{L}^\infty  dz' \int_0^{2\pi}
d\phi'\,\frac{r
a\,\sin\theta\cos^2\phi'}{g(r,\theta,z')^{3/2}}\right|
\]\[\leq
\mathrm{cte}\int_{L}^{\infty}dz'\,\frac{1}{(z'^2 - 2 r z')^{3/2}}\leq
\mathrm{cte}\int_{L}^{\infty}dz'\,\frac{1}{z'^3}=\frac{\mathrm{cte}}{L^{2}},
\]which vanishes as $L\to\infty$. Note that we have got an upper bound to
the rate of convergence as $L^{-2}$ (this
rate was also found numerically).

Now we check that the above expressions for $A_{L,\phi}(r,\theta)$
actually result in the right gauge in the limit
$L\to\infty$, that is, in cylindrical coordinates $\rho=r\sin\theta$,
\[
A_{\phi}(r,\theta)= \left\{ {\matrix{\Phi/(2\pi \rho)\quad \rho\ge a>0\cr
\Phi \rho/(2\pi a^2)\quad 0\le \rho\le a \cr
}} \right.
.
\] For this it is enough to calculate the vector potential for some range
of $r,\theta$, say $r\gg a$ and $r\ll a$.

Let's consider the case of a point $\mathbf x$ far from the solenoid, that
is, $r\sin\theta\gg a$.
Substitute the above expression for $\cos\phi'/f(r,\theta,z',\phi')$ in
$A_{L,\phi}(r,\theta)$ so that, after performing
the resulting integrals,
$$A_{L,\phi}(r,\theta) \approx
\frac{\Phi}{2\pi}\frac{r\sin\theta}{r^2+a^2-r^2\cos^2\theta} \frac{\alpha
(L-r\cos\theta)+\beta(L+r\cos\theta)}{2\beta\alpha},$$
with $\alpha = \sqrt{r^2+a^2+L^2+2 r\cos\theta L}$ and  $\beta =
\sqrt{r^2+a^2+L^2-2 r\cos\theta L}$. Hence, for large $L$
$$A_{L,\phi}(r,\theta) \approx
\frac{\Phi}{2\pi}\frac{r\sin\theta}{r^2+a^2-r^2\cos^2\theta}.$$
Taking into account that $r\sin\theta\gg a$ again, we see that
$A_{L,\phi}$ approaches
$A_\phi$ above as $L\rightarrow\infty$, and the right gauge is obtained.
Similar arguments hold for $r\sin\theta\ll a$.
Observe that for $\theta=\pi/2$ the above steps infer the convergence in
the $xy$ plane, that is, in $\R^2$.

We underline that for points $\mathbf x\notin \mathcal S$ the integrand in
the expression for  $A_{L,\phi}$ is a
continuous function and, for fixed $r,\theta$, with $r\sin\theta\ne a$,
there is $d>0$ so that  the absolute value of
the denominator in the integrand is uniformly $\ge d$. In fact, one can take
$$\displaystyle d\ddd\min_{\mathbf{x}'\in\mathcal
S}|\mathbf{x}-\mathbf{x}'|= |r\sin\theta - a| > 0.$$ In summary, off the
solenoid border, the above expressions for
the vector potentials result in finite values for both $L<\infty$ and
$L=\infty$.

For points $\mathbf{x}$ on the solenoid border, that is,
$|\mathbf{x}-\mathbf{x}'| = 0$, for some $\mathbf{x}' \in \mathcal S$,
the denominator of  the integrand in the expression for  $A_{L,\phi}$
vanishes, which causes a divergence in the
integrals; however, such expression for $A_{L,\phi}$ is not  supposed to
hold on this  border, and the values of
$\mathbf A$ are recovered by continuity (by using lateral limits from
inside and outside of the solenoid).
In any event, the solenoid border is a set of zero Lebesgue measure in
$\R^3$ and $\R^2$.

Finally, note that it is not necessary to consider a finite solenoid with
$-L<z'<L$, since all arguments are easily
adapted to $-L_1<z'<L_2$, with $L_1\to\infty,L_2\to\infty$.

\

 \subsubsection*{Acknowledgments} {\small The authors acknowledge partial
support from CNPq (Brazil).}

\

\end{document}